\documentclass[reprint,secnumarabic,amssymb,amsmath,aip,cha,groupedaddress,frontmatterverbose]{revtex4-1}
\usepackage{graphicx}

\usepackage{amsmath}
\usepackage{booktabs}
\usepackage{color}

\begin{document}

\title{Spectral characteristics of  time resolved magnonic spin Seebeck effect}

\author{S. R. Etesami, L. Chotorlishvili, and J. Berakdar}
\affiliation{Institut f\"ur Physik, Martin-Luther-Universit\"at Halle-Wittenberg, 06099 Halle, Germany}

\begin{abstract}
 Spin Seebeck effect (SSE) holds promise for new spintronic devices with low-energy consumption. The underlying physics, essential for a further progress, is yet to be fully clarified.  This study of the time resolved longitudinal SSE in the magnetic insulator yttrium iron garnet (YIG) concludes that a substantial  contribution  to the spin current stems from small wave-vector subthermal exchange magnons. Our finding is in line with the recent experiment by  S. R. Boona and J. P. Heremans, Phys. Rev. B \textbf{90}, 064421 (2014). Technically, the spin-current dynamics is treated  based on the  Landau-Lifshitz-Gilbert (LLG) equation also including    magnons back-action on
thermal bath, while the formation of the time dependent thermal gradient is described self-consistently via the heat equation coupled to the magnetization dynamics.

\end{abstract}
\date{\today}

\maketitle

\textbf{\textit{Introduction.}} The spin counterpart of the Seebeck effect, the  spin-Seebeck effect (SSE) refers to the emergence of a spin current upon  applying a thermal bias. Aside  fundamental interest,  SSE is of relevance for a variety of applications, including spin-dependent
thermoelectric devices \cite{Kirihara}. Since its first observation \cite{Uchida2008}, SSE  has been studied extensively  both experimentally \cite{Jaworski,Uchida2010,Kajiwara,Agrawal,Agrawal2,Boona,TRSSE,Uchida2014,Lee,Qiu,Qiu2,Brandon,Qiu3} and theoretically \cite{Ohe,Chotorlishvili,Etesami,Xiao,Adachi,Hoffman,Schreier,Toelle} and for a variety of systems. In particular,  SSE in ferromagnetic (FM) insulators  \cite{Uchida2010}  hints on a magnonic origin of the spin current \cite{Kajiwara}. Magnonic SSE has some advantages with respect to charge-carrier-related spin current  in that magnon spin current propagates over a length scale of up to a  millimeter \cite{Adachi2} while conduction-based spin current is usually much smaller due to spin-dependent scattering.  The spectral characteristics of the magnons contributing to the magnonic SSE  are still  under debate.
In a recent spatially resolved experiment \cite{Agrawal} on the magnetic insulator yttrium iron garnet (YIG), the measured magnon temperature $T_{m}$  was related to the short wavelength exchange part of the magnon spectrum $\omega \big(\vec{k}\big)$. An important observation is  that a non-vanishing spin current emerges even for equal magnon and phonon temperatures $T_{m}=T_{ph}$. Note that the standard narrative attributes the emergence of the magnonic spin current to the difference between magnon and phonon temperatures. A theoretical explanation to the observed fact was given in terms of the long-wavelength dipolar part of the magnon spectrum \cite{Rückriegel}. If long-wavelength dipolar magnons are weakly coupled to the phonons their life time is larger and the magnon temperature   deviates from the phonon temperature. Thus,  dipolar magnons might contribute to SSE. This explanation  though comprehensible,  does not exclude the contribution of the long-wavelength exchange subthermal magnons in the SSE. As was shown in a recent theoretical paper \cite{Etesami} constitutive issue in the formation of the magnonic SSE is not the difference between magnon and phonon temperatures but rather the nonuniform magnon temperature profile leading to a nonzero exchange spin torque and the magnon accumulation effect that drives the magnonic SSE. In the present paper we  study the contribution of the long-wavelength (small wave-vector $\vec{k}$ ) exchange subthermal magnons to the magnonic SSE. We find that contrary to the short-wavelength exchange thermal magnons, the long-wavelength exchange subthermal magnons do  contribute substantially  to the SSE. Our predictions  are consistent with  the recent experimental report  by S. R. Boona and J. P. Heremans \cite{Boona}. Magnon modes at thermal energies in YIG were found not responsible for the spin Seebeck effect. Subthermal magnons, i.e., those at energies below about $30\pm10$ [K], were found important for the spin transport in YIG at all temperatures. To assess the partial contribution to SSE of  magnons with different frequencies  $\omega \big(\vec{k}\big)$  we analyze the time resolved SSE  adopting a low-pass filter, as done experimentally  \cite{TRSSE}. In time resolved SSE experiments  the laser modulation frequency is the relevant control parameter. If it is smaller compared to the cutoff frequency of the filter then the filter does not affect the spin current, otherwise the low-pass filter cuts the spin current (external cutoff).  Any cutoff observed for modulation frequencies smaller than cutoff frequency of the filter is thus intrinsic. In our simulation we implemented this approach, which aside from mimicking the experimental situation bears some advantages against a discrete Fourier transform to obtain the spectral dependence of the spin current, as detailed in the supplementary materials \cite{supp}.  \\
 We note, that while our study is limited to thermally induced magnonic transport in FM insulators,  it can be in principle extended to  include
 contributions from thermally activated carriers in metals or semiconductors. The method would rely however on
 further  reasonable inputs such as the  details of the spin torque current and the rescaled exchange interaction parameters.

\begin{center}
   \begin{figure}[!t]
    \centering
    \includegraphics[width=\columnwidth]{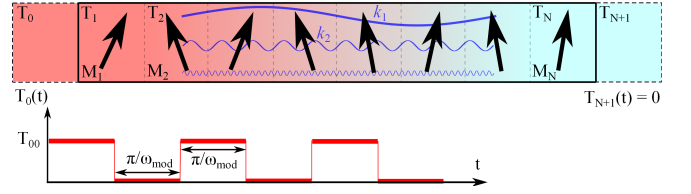}
        \caption{\label{skim} Schematic of the contribution of the spin waves with different wave vectors $k_i$ to the spin current in SSE.
         The temperature of the left edge of the system is varied in time: $T_0(t)=T_{00}S\left(\omega_\mathrm{mod}t\right)$, where $S\left(\omega_\mathrm{mod}t\right)$ is a rectangular pulse with a modulation frequency  $\omega_\mathrm{mod}$ and levels $0$ and $1$. $T_n$ and $M_n$ represent the temperature and magnetization in each individual cell, respectively, and  are calculated self-consistently via the heat and LLG equations (Eqs.~(\ref{FTCS}) and (\ref{LLG_discrete})).}
    \end{figure}
\end{center}

\textbf{\textit{Theoretical background}}. In a FM insulator, the low-lying excitations are spin waves describable  by Landau-Lifshitz-Gilbert (LLG) equation (for a comparison an implementation based on the  Landau-Lifshitz-Miyasaki-Seki  scheme  has also been performed
with full details and results being included in the Supplementary Materials \cite{supp})
\begin{equation}
\label{LLG}
\begin{split}
\frac{\partial}{\partial t}\vec{M}(\vec{r},t)=&-\gamma\vec{M}(\vec{r},t)\times\left[H_0\hat{z}+\frac{2A}{M_S^2}\nabla^2\vec{M}(\vec{r},t)\right]\\
&+\frac{\alpha}{M_S}\vec{M}(\vec{r},t)\times\frac{\partial}{\partial t}\vec{M}(\vec{r},t),
\end{split}
\end{equation}
where $\vec{M}(\vec{r},t)$ stands for the magnetization vector, $\gamma$ is the gyromagnetic ratio, $H_0$ is the external magnetic field, $A$ is the exchange stiffness, $M_S$ is the saturation magnetization and $\alpha$ refers to the Gilbert damping constant. The LLG equation in the linear limit has the following solution $M_x(\vec{r},t)+iM_y(\vec{r},t)\propto \exp\left(i\vec{k}\cdot\vec{r}+i\omega_kt\right)\exp\left(-\alpha\omega_kt\right)$. The spin wave dispersion relation $\omega_k=\gamma\left(H_0+\frac{2A}{M_S}k^2\right)$ and the damping which depends on the wave vector $\alpha\omega_k$, \cite{Kajiwara} indicates that short wave vector $\vec{k}$ (long wavelength) exchange magnons are less damped. Thus magnons have different relaxation times. Assuming that the magnon-phonon scattering is the main source of damping, the magnon-phonon relaxation (thermalization) time reads
\begin{equation}
\label{magnon-phonon-relaxation-time}
\tau_{mp}^k=\left[\alpha\gamma\left(H_0+\frac{2A}{M_S}k^2\right)\right]^{-1},k=2n\pi/d,n=0,\pm1,\cdots
\end{equation}
where $d$ is the length of the FM insulator.
Let us  assume that the left edge of the FM insulator  is heated up periodically   with the modulation frequency $\omega_{\mathrm{mod}}$ (see Fig.~\ref{skim}). A time periodic thermal bias is meant to mimic qualitatively  the action of  laser pulses
 (in real experiment the laser electromagnetic energy is absorbed by the sample only partly, however. More details are found in the
 Supplementary Materials\cite{supp}).
 The inherent thermal loses relevant to the experiment are treated self-consistently  by adopting  an additional source term  in the heat equation (see Eq.~(SM-3) in the Supplementary materials\cite{supp}). Thus, the temperature profile implemented in the LLG equation in our case is calculated self-consistently.
 For unraveling the role of magnons with different frequencies we follow the idea
of using a low-pass filter, i.e., a filter with an extrinsic cutoff frequency $\omega_{c}$ which detruncates the
spin current if the modulation frequency of the laser pulses exceeds the extrinsic cutoff frequency of the filter $\omega_{mod}>\omega_{c}$.
The ratio between "true" $I_{0}$ and measured $I\big(\omega_{mod}\big)$ spin current reads $I\big(\omega_{mod}\big)=I_{0}/\sqrt{1+\big(\omega_{mod}/\omega_{c}\big)^{2}}$. If $\omega_{mod}<\omega_{c}$, the
measured spin current is not altered by the filter. The decay in the spin current in this case is ascribed to the intrinsic cutoff frequencies, which in turn are related to the magnon-phonon-relaxation-time (see Eq.~(\ref{magnon-phonon-relaxation-time})) $\Omega_{mp}^k=2\pi/\tau_{mp}^k$. In this way  different internal cutoff frequencies can  be observed.

\textbf{\textit{Model and simulations.}} We model a ferromagnetic insulator via a chain of FM cells arranged  along the  $\hat{x}$ axis (Fig.~\ref{skim}). The total energy density of the system of  $N$ cells reads:
\begin{equation}
\label{energy_density}
e=-H_0\sum_{n=1}^{N}M_n^z-\frac{2A}{a^2M_S^2}\sum_{n=1}^{N}\vec{M}_n\cdot\vec{M}_{n+1},
\end{equation}
where $a$ is the size of the cell, $\vec{M}_n$ is the magnetization vector of $n^{th}$ cubic cell and $A$ is the exchange stiffness. We use Eq.~(\ref{energy_density}) to model Yttrium-iron-garnet (YIG) which has been employed extensively in  SSE experiments.
 The effective magnetic field acting on the  $n^{th}$ cell reads $\vec{H}_n^\mathrm{eff}(t)=-\frac{\partial e}{\partial \vec{M}_n}=H_0\hat{z}+\frac{2A}{a^2M_S^2}\left(\vec{M}_{n+1}+\vec{M}_{n-1}\right)$.

A Gaussian-white noise $\vec{\eta}_n(t)$ contribution to the effective magnetic field (with a correlation function $\langle\eta_{ni}(t)\eta_{mj}(t+\Delta t)\rangle=\frac{2\alpha K_B T_n(t)}{\gamma M_S a^3}\delta_{nm}\delta_{ij}\delta(\Delta t)$) accounts for thermal fluctuations. Here, $\langle\cdots\rangle$ means average over different realization of the noise, $n$ and $m$ are cell numbers and $i$ and $j$ are the Cartesian components. The time and site-dependent temperature $T_n(t)$ obeys the following heat equation
\begin{equation}
\label{FTCS}
\frac{d}{d t}T_n(t)=\frac{\kappa}{\rho C}\frac{T_{n+1}(t)-2T_{n}(t)+T_{n+1}(t)}{a^2},
\end{equation}
with the initial and the boundary conditions
\begin{equation}
\label{temperature_boundary}
\begin{split}
&T_n(t=0)=0;\qquad n=0,\cdots N+1\\
&T_0(t)=T_{00}S\left(\omega_\mathrm{mod}t\right)\qquad,T_{N+1}(t)=0.\\
\end{split}
\end{equation}
$\kappa$ is the phononic thermal conductivity, $\rho$ is the mass density, $C$ is the phonon heat capacity, $T_{00}$ is the temperature applied on the left edge and $S$ is a series of rectangular laser pulses with the modulation frequency  $\omega_\mathrm{mod}$  (see Fig.~\ref{skim}). For solving the heat equation we implemented a Forward-Time Central-Space (FTCS) scheme \cite{Press}. The hierarchy of the relaxation times for phonons $\tau_{ph}=\left[\frac{\kappa}{\rho C}\left(\frac{\pi}{Na}\right)^2\right]^{-1}\approx 10 [\textmd{ns}]$ (the number and size of cells $N=50$, $a=10 [\textmd{nm}]$ and phonon thermal conductivity $\kappa=6 [\textmd{W}.\textmd{m}^{-1}.\textmd{K}^{-1}]$) and magnons $\tau_{mp}=\frac{1}{2\alpha \omega_{0}}\approx 10^{3}[\textmd{ns}]$ (ferromagnetic resonance frequency and phenomenological damping constant $\omega_{0}=10$ [Ghz], $\alpha\approx10^{-4}$) allows an adiabatic decoupling which amounts, to a first order, to plug the obtained phonon temperature profile directly in the LLG equation and study so the magnetization dynamics self-consistently.

\begin{center}
   \begin{figure}[!t]
    \centering
    \includegraphics[width=\columnwidth]{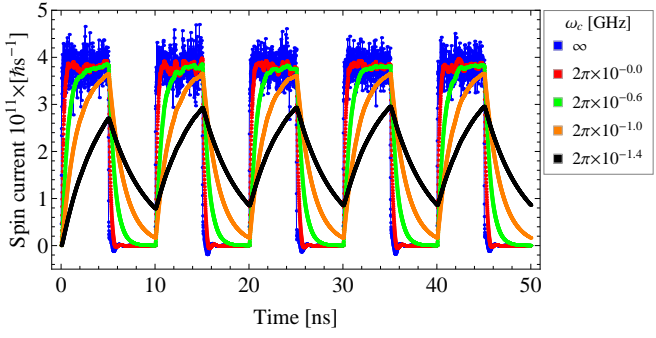}
        \caption{\label{spin_current_vs_time} Spin current at the middle of a chain of $50$ FM cells versus time and for different extrinsic cutoff frequencies ($\omega_c$). We choose
         $a=10[\textmd{nm}]$, $H_0=0.057[\textmd{T}]$, $T_{00}=10[\textmd{K}]$, $\alpha=0.1$. The system is heated up periodically with the modulation frequencies of $\omega_{\mathrm{mod}}=2\pi\times10^{-1.0}[\textmd{GHz}]$ and the spin current is statistically averaged over $1000$ realization of the noise. For the blue curve no cutoff frequency is implemented on the spin current ($\omega_c=\infty$) but for the red, green, orange and black curves the cutoff frequencies are $\omega_c=2\pi\times10^{0.0}[\textmd{GHz}]$, $\omega_c=2\pi\times10^{-0.6}[\textmd{GHz}]$, $\omega_c=2\pi\times10^{-1.0}[\textmd{GHz}]$ and $\omega_c=2\pi\times10^{-1.4}[\textmd{GHz}]$, respectively.}
    \end{figure}
\end{center}

\begin{center}
   \begin{figure}[!t]
    \centering
    \includegraphics[width=\columnwidth]{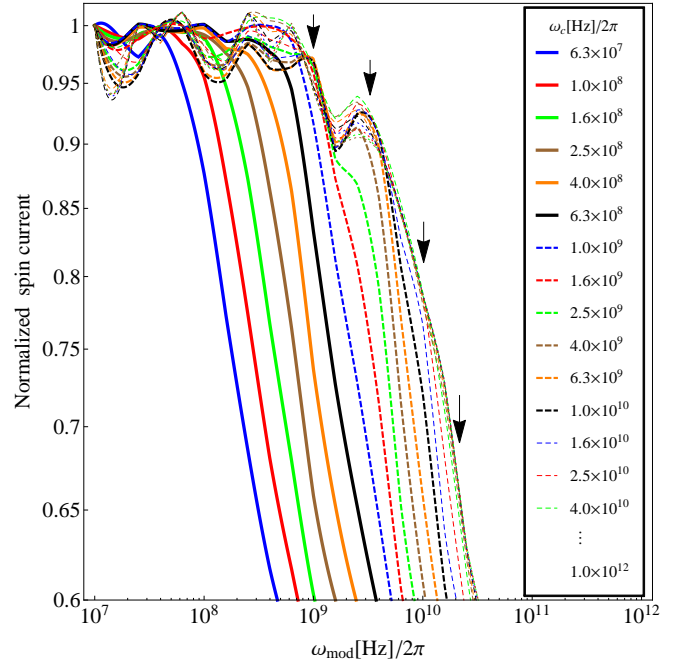}
        \caption{\label{spin_current_vs_omega} Normalized spin current ($\frac{I_{\omega_{mod}}}{I_{2\pi\times10^7[\textmd{Hz}]}}$) at the middle of a chain of $50$ FM cells versus the modulation frequency for different extrinsic cutoff frequencies ($\omega_{c}$) with the parameters $a=10[\textmd{nm}]$, $H_0=0.057[\textmd{T}]$, $T_{00}=10[\textmd{K}]$, $\alpha=0.1$. The spin current is statistically averaged over $1000$ noise realizations. For $\omega_{c}<2\pi\times10^{9}$[\textmd{Hz}] the cascades follow the extrinsic cutoff frequencies which are characteristic of Low-Pass filter. However, for $\omega_{c}\geq2\pi\times10^{9}$[\textmd{Hz}] the cascades occur earlier than the corresponding extrinsic cutoff which are a sign of inherent intrinsic cutoff in the system. The arrows show the magnon-phonon frequencies ($\Omega_{mp}^k/2\pi$ in TABLE~\ref{tab_1}) for different wave vectors evaluated  theoretically  (Eq.~(\ref{magnon-phonon-relaxation-time})) and coinciding with the appearance of intrinsic cutoff frequencies (cascades) in the curves.}
    \end{figure}
\end{center}

The magnetization dynamics is governed by a set of coupled LLG equations
\begin{equation}
\label{LLG_discrete}
\begin{split}
&\frac{\partial}{\partial t}\vec{M}_n(t)=\\
&-\frac{\gamma}{1+\alpha^2}\vec{M}_n(t)\times\left[\vec{H}_n^\mathrm{eff}(t)+\frac{\alpha}{M_S}\vec{M}_n(t)\times\vec{H}_n^\mathrm{eff}(t)\right].\\
\end{split}
\end{equation}
For the numerical integration of the coupled stochastic differential equations we utilize the Heun's method \cite{Kampen,Kloeden}.
\begin{table}[!b]
\caption{Magnon-phonon relaxation times and the corresponding frequencies according to Eq.~(\ref{magnon-phonon-relaxation-time}) for $N=50$, $a=10 [\textmd{nm}]$, $H_0=0.057 [\textmd{T}]$ and $\alpha=0.1$.}
\centering
\vspace{2.ex}
\begin{tabular}{|c|c|c|c|c|c|c|}
\hline
 $n$ & 0 & 1 & 2 & 3 & 4 & 5\\ \hline
 $|k|=2\pi n/Na$ [$10^8m^{-1}$] & 0.00 & 0.13 & 0.25 & 0.38 & 0.50 & 0.63\\ \hline
 $\tau_{mp}^k$ [ns] & 1.000 & 0.303 & 0.098 & 0.046 & 0.026 & 0.017\\ \hline
 $\Omega_{mp}^k/2\pi$ [GHz] & 1.0 & 3.3 & 10.2 & 21.6 & 37.7 & 58.4\\ \hline
\end{tabular}
\label{tab_1}
\end{table}
The spin current tensor is calculated using the following formula: $I^{\alpha}_{n}=-\frac{2Aa}{M_{S}^{2}}\sum_{m=1}^{n}\langle M^{\mathrm{\beta}}_{\mathrm{m}}(M^{\mathrm{\gamma}}_{\mathrm{m-1}}+M^{\mathrm{\gamma}}_{\mathrm{m+1}})\rangle\varepsilon_{\alpha\beta\gamma}$ where $\alpha=x,y,z$ defines the spin components of the  tensor, while $n$ stands for the cite number, $\varepsilon_{\alpha\beta\gamma}$ is the Levi-Civita antisymmetric tensor and $\langle\cdots\rangle$ means averaging over the different realization of the noise \cite{Kajiwara,Etesami,Ohe,Hoffman}. In our model, because of the particular geometry of the system (1D chain aligned along $\hat{x}$ axis) the only non-zero element of the spin current tensor is $I^{z}_{n}$ \cite{Etesami}.
For the output signal we implemented a recursive low-pass filter \cite{Press}
$I_{out}(t)=\frac{\omega_c\Delta t}{1+\omega_c\Delta t}I_{in}(t)+\frac{1}{1+\omega_c\Delta t}I_{out}(t-\Delta t)$. Here $I_{in}$ and $I_{out}$ are the spin currents before and after filtering procedure and $\omega_c$ is the extrinsic cutoff frequency of the filter (see Supplementary Materials\cite{supp}).

\begin{center}
   \begin{figure}[h]
    \centering$
        \begin{array}{cc}
    \includegraphics[width=\columnwidth]{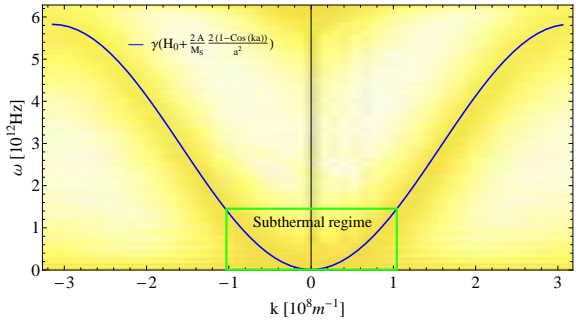}\\
        \end{array}$
        \caption{\label{SWdispersion} Spin-wave dispersion relation: The yellow background shows the absolute value of the discrete fourier transformation of $m_x+im_y$ based on micromagnetic simulations \cite{Kumar} for a chain of $50$ FM cells under a linear temperature gradient. $T_1=0$, $T_N=10$ [K], $a=10[\textmd{nm}]$, $H_0=0.057[\textmd{T}]$. For small $k$, the dispersion relation reduces to $\omega_k=\gamma\left(H_0+\frac{2A}{M_S}k^2\right)$. The small green frame shows the subthermal regime of the spectrum.}
    \end{figure}
\end{center}

\textbf{\textit{Results and discussion.}}   Fig.~\ref{spin_current_vs_time}  shows the spin Seebeck current as a function of time for different extrinsic cutoff frequencies $\omega_{c}$. To each cutoff frequency a certain color is attributed. The filter with the extrinsic cutoff frequency $\omega_{c}$ cuts the spin current if the modulation frequency of the laser pulses $\omega_{mod}$ exceeds the extrinsic cutoff $\omega_{mod}>\omega_{c}$.
Thus, the larger  the extrinsic cutoff of the filter $\omega_{c}$, the  larger is the spin current. This is what we see in  Fig.~\ref{spin_current_vs_time}.
For  small extrinsic cutoff frequencies (see Fig.~\ref{spin_current_vs_omega})  $\omega_{c}<2\pi\times10^{9}$[\textmd{Hz}] the spin current is detruncated extrinsically at  modulation frequencies $\omega_{mod}=\omega_{c}$ smaller than the first intrinsic cutoff frequency of the system $\Omega_{mp}^{k=0}=2\pi/\tau_{mp}^{k=0}$. Therefore, no inherent cutoff is observed in this case. However, for an elevated extrinsic cutoff $\omega_{c}$ we observe a cascade of the inherent intrinsic cutoff at the frequencies $\Omega_{mp}^k=2\pi/\tau_{mp}^k<\omega_{c}$. All these intrinsic cutoff frequencies are in the subthermal regime of the magnon spectrum $k<k_{max}\approx10^8$ [$m^{-1}$] \cite{Agrawal,Boona} (TABLE~\ref{tab_1}). To be confident while heating up the system the thermal magnons are also activated, the corresponding dispersion relation to our parameters is shown in Fig.~\ref{SWdispersion}. As can be seen, magnons with a broad range of frequencies, beyond subthermal regime are also activated. The subthermal regime of the spectrum is shown with a small green frame\cite{Agrawal,Boona}.

\begin{center}
   \begin{figure}[b!]
    \centering$
        \begin{array}{ccc}
    \includegraphics[width=7cm,height=16cm]{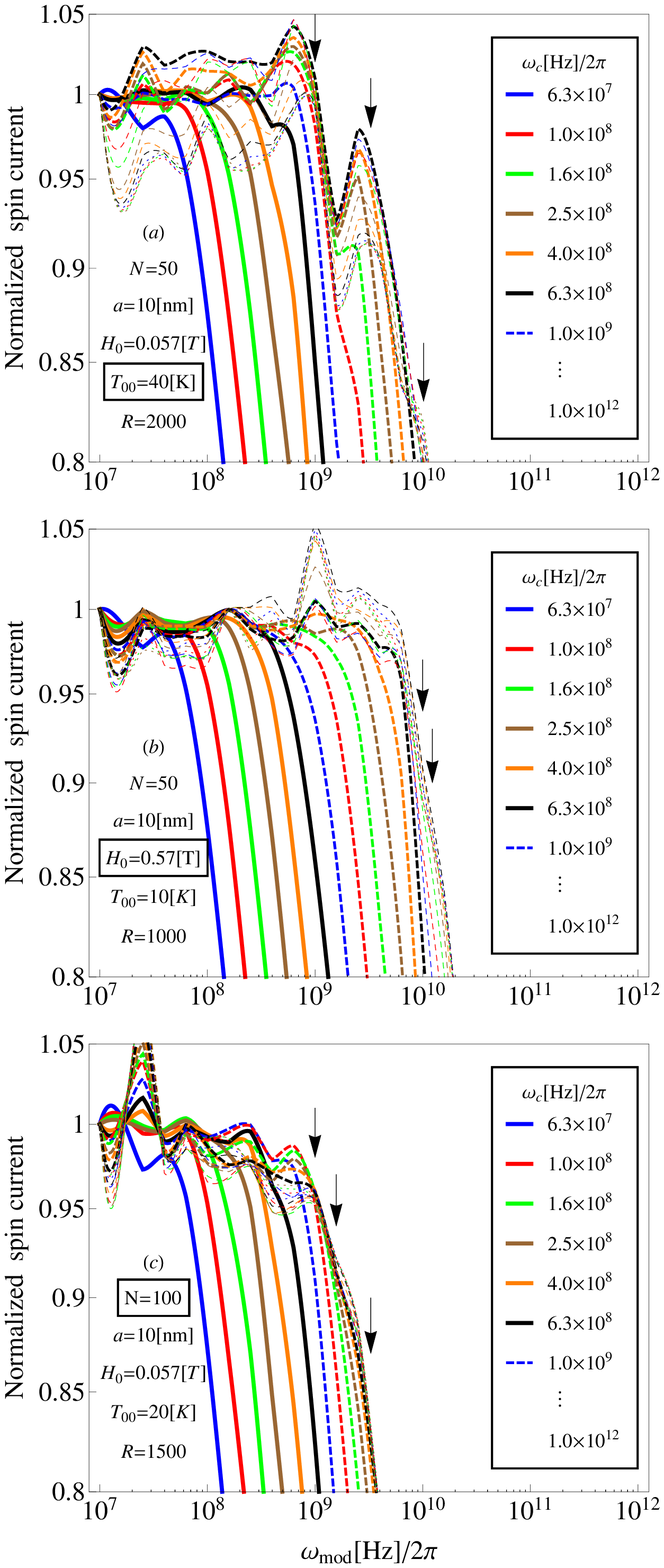}
        \end{array}$
        \caption{\label{spin_current_vs_omega_d} Normalized spin current ($\frac{I_{\omega_{mod}}}{I_{2\pi\times10^7[\textmd{Hz}]}}$) at the middle of a FM chain versus modulation frequency for different extrinsic cutoff frequencies ($\omega_{c}$). $R$ is the number of realization of the noise over which the spin current is statistically averaged. The arrows show the magnon-phonon frequencies ($\Omega_{mp}^k/2\pi$) for different wave vectors calculated theoretically(Eq.~(\ref{magnon-phonon-relaxation-time}))  and compared to the appearance of  the intrinsic cutoff frequencies (cascades) in the curves.}
    \end{figure}
\end{center}

To ensure that our findings are not an artefact of a particular choice of parameters/model we performed the calculations for  various temperature gradients, different applied external magnetic fields, and varying lengths of the chain (Fig.~\ref{spin_current_vs_omega_d}). In all these cases the intrinsic cutoff frequencies follow the corresponding magnon-phonon frequencies (Eq.~(\ref{magnon-phonon-relaxation-time})). Furthermore, the use of a white noise for a swift time-dependent heating might be questioned.
Therefore, we implemented the  Landau-Lifshitz-Miyasaki-Seki  scheme \cite{Miyazaki}
to account for the back-action of the magnon subsystem to the surrounding
phonon bath (see Supplementary Materials\cite{supp}) and arrived  basically at the same conclusion that
small-wave vectors $\vec{k}$ exchange subthermal magnons
 contribute substantially to the formation of thermally activated spin current. Hence, our finings are expected to be of some generalities
  for magnon-driven SSE and associated devices, for the employed schemes are quite ubiquitous,   had proven
 to be reliable for finite temperature spin dynamics, and the employed system parameters are generic.

\textbf{\textit{Acknowledgements.}} We thank K. Zakeri Lori, Y.-J. Chen and  A. Sukhov for valuable discussions.

\end{document}